\documentclass[aps,prd,preprintnumbers,showpacs]{revtex4}
\usepackage[dvips]{graphicx}
\begin{document}

%
%

\eprint{Nisho-3-2014}
\title{Fast Radio Bursts from Axion Stars}
\author{Aiichi Iwazaki}
\affiliation{International Economics and Politics, Nishogakusha University,\\ 
6-16 3-bantyo Chiyoda Tokyo 102-8336, Japan.}   
\date{Dec. 25, 2014}
\begin{abstract}
Axions are one of the most promising candidates of dark matter.
The axions have been shown to form miniclusters with masses $\sim 10^{-12}M_{\odot}$
and to become dominant component of dark matter.
Some of the axion miniclusters condense to form axion stars. 
We have recently shown a possible origin of fast radio bursts ( FRBs ) 
by assuming that the axion stars are main component of halos: FRBs
arise from the collisions between the axion stars and neutron
stars. 
It is remarkable that the masses of the axion stars obtained by the comparison
of the theoretical and observational event rates are coincident with
the mass $\sim 10^{-12}M_{\odot}$. 
In this paper, we describe our model of FRBs in detail.
We derive the approximate solutions of the axion stars with large radii
and constraint their masses for the approximation to be valid.
The FRBs are emitted by electrons in atmospheres of neutron stars.
By calculating the optical depth of the atmospheres,
we show that they are transparent for the radiations with
the frequency given by the axion mass $m_a$ such as $m_a/2\pi\simeq 2.4$GHz$(m_a/10^{-5}\rm eV)$.
Although the radiations are linearly polarized when they are emitted,
they are shown to be circularly polarized after they pass magnetospheres of neutron stars. 
We also show that the FRBs are not broadband and their frequencies have finite bandwidths 
owing to the thermal fluctuations of the electrons in the atmospheres of the neutron stars.
The presence of the finite bandwidths is a distinctive feature of our model
and can be tested observationally.
Furthermore, we show that similar FRBs may arise 
when the axion stars collide with magnetic white dwarfs $B\sim 10^9$G.
The distinctive feature is that
the durations of the bursts are of the order of $0.1$second
and that the radiations have wider bandwidths than 
those of the radiations produced by the collisions with neutron stars.
\end{abstract}
\hspace*{0.3cm}
\pacs{98.70.-f, 98.70.Dk, 14.80.Va, 11.27.+d \\
Axion, Neutron Star, Fast Radio Burst}

\hspace*{1cm}

\maketitle

\section{introduction}
Fast Radio Bursts have recently been discovered\cite{frb,frb1,frb2} at around $1.4$ GHz frequency.
The durations of the bursts are typically
a few milliseconds. The origin of the bursts has been suggested to be
extra-galactic owing to their large dispersion measures.
This suggests that the large amount of the energies $\sim 10^{43}$GeV/s
is produced at the radio frequencies.
The event rate of the burst is estimated to be $\sim 10^{-3}$ per year in a galaxy.
Furthermore, no gamma or X ray bursts associated with the bursts
have been detected. Follow up observations\cite{frb3} of FRBs do not find any 
signals from the direction of the FRB. 
To find progenitors of the bursts, several models\cite{model} have been proposed.
They ascribe FRBs to traditional sources such as neutron star-neutron star mergers, 
magnetors, black holes, et al.. 

Our model\cite{i,t} ascribes FRBs to axions\cite{axion}, 
which are one of most promising candidates of
dark matter. A prominent feature of axions is that they are converted to 
radiations under strong magnetic fields. 
The axions form axion stars
known as oscillaton\cite{axion2} made of axions bounded gravitationally.
The axion stars are condensed objects 
of axion miniclusters\cite{kolb}, which
have been shown to be produced after the QCD phase transition and to 
form the dominant component of dark matter in the Universe. 
Furthermore,
the axion miniclusters
have been shown to form the axion stars by gravitationally losing their kinetic energies\cite{kolb,axions}.
Thus, the axion stars are the dominant component of dark matter. 
We have recently proposed a progenitor of the FRBs
that the FRBs arise from the collisions between the axion stars and neutron stars.  
All of the properties ( duration, event rate and total radiation energy ) 
of the FRBs observed can be naturally explained in our model.

In the paper, we present the details of our models. First of all, we derive
the solutions of the axion stars by expanding
the axion field and the gravitational fields 
in terms of eigen modes $\cos(n\omega t)$ with $n$ integers,
where the eigen value $\omega$ can be determined 
by solving axion field equation representing gravitationally bounded axions. It is given such that
$\omega=m_a-k^2/2m_a$ where $k^2/2m_a$( $\ll m_a$ ) 
denotes a gravitational binding energy of an axion bounded by
the axion star; $k\simeq GM_am_a^2$ where $G$ and $M_a$ are the gravitational constant and the mass of
the axion star, respectively.
Then, we find that the mass $M_a$ of the axion star is given 
in terms of the radius $R_a=1/k$ of the axion stars 
such that $M_a=1/(Gm_a^2R_a)$.
The value of the mass $M_a$ can be obtained by the comparison of the theoretical and observational 
even rate of FRBs in our production mechanism of the FRBs. 
It is remarkable that a mass $M_a\sim 10^{-12}M_{\odot}$ obtained 
in such a way is
coincident with the mass of the axion miniclusters previously obtained. 
The radus $R_a$ is given by $R_a\sim 10^2$km,
which is much larger than neutron stars.
 
The field configurations of the solutions
are not static but oscillating and localized with radius $R_a=1/k$; 
$a(t,\vec{x})\propto \cos(\omega t)\exp(-k|\vec{x}|) \simeq\cos(m_at)\exp(-k|\vec{x}|)$.
We show that the oscillating electric fields 
$\vec{E}_a(t,\vec{x})\propto a(t,\vec{x})\vec{B}(\vec{x})$ are
generated on the axion stars under external magnetic fields $\vec{B}(\vec{x})$. 
Thus, when the axion stars collide with neutron stars with strong magnetic fields,
the electric fields $\vec{E}_a(t,\vec{x})$ are generated, which make electrons in atmospheres\cite{atmos} of 
neutron stars coherently oscillate.
Thus, the electrons emit coherent radiations with the frequency given by the axion mass. 
Since the electrons are much dense in the atmospheres, 
the large amount of radiations with the frequency 
$m_a/2\pi\simeq 2.4\,\mbox{GHz}\,(m_a/10^{-5}\mbox{eV})$
can be produced in the collisions. 
The total amount of the energy of the radiations is given by  
$10^{-12}M_{\odot}(10\rm km/10^2\rm km)^2\sim 10^{43}$GeV,
where the radii of the neutron stars and the axion stars are supposed to be $10$km and $10^2$km, respectively.
This is our production mechanism of FRBs.



We also discuss the optical properties of the atmospheres of neutron stars.
The geometrical depth of the hydrogen atmospheres of old neutron stars is of the order of $0.1$cm.
The radiations emitted in the atmospheres can pass through them because 
they are shown to be optically thin for the radiations with the frequency
$(m_a/2\pi)\simeq 2.4 \,\mbox{GHz}\,(m_a/10^{-5}\mbox{eV})$.
They can also pass through magnetospheres of neutron stars.
After they pass the magnetospheres, the radiations 
are circularly polarized owing to the absorption of 
the radiations with either right or left handed polarization. 
The circular polarization of a FRB has recently been observed\cite{frb3}.

Although the radiations produced by our mechanism 
are monochromatic having the frequency given by $\omega=m_a/2\pi$,
the observed radiations have bandwidths at least wider than the range $1.2$GHz$\sim 1.6$GHz
used by actual observations. We show that
the bandwidths $\omega_{\rm th}$ ( $\omega \pm \omega_{\rm th}$ ) 
of FRBs arise from thermal fluctuations of electrons. 
Thus, the bandwidths are narrow.
The presence of such narrow bandwidths owing to the thermal 
fluctuations is a distinctive feature of our
model and can be tested observationally.

The relative velocities $v_c$ in the collisions 
are given by $\sqrt{2GM_{ns}/R_{ns}}$,
which is of the order of $10^5$km/s; $M_{ns}$ ( $R_{ns}$ ) denote mass ( radius ) of
neutron stars.
The fact explains the duration of the FRBs being of the order 
of $\sim 1$ms. The radiations observed at the rest frame of neutron stars
are affected by Doppler effect. Furthermore, the radiations observed at the earth
are affected by gravitational ( $\sqrt{1-2GM_{ns}/R_{ns}}$ ) 
and cosmolgical redshifts $z$. Owing to the effects,
the actual frequencies observed at the earth are given by 
$(m_a/2\pi)\times (1-2GM_{ns}/R_{ns})/(1+z)$ which is less than 
$(m_a/2\pi)\simeq 2.4 \,\mbox{GHz}\,(m_a/10^{-5}\mbox{eV})$.

\vspace{0.1cm}

It is interesting to see that similar radio bursts are emitted
when the axion stars collide with white dwarfs with very strong 
magnetic fields $\sim 10^9$G.
In particular,
the duration of the bursts is of the order of $0.1$second, contrary to
the observed FRBs. This is because the relative velocities between the axion stars and the white dwarfs
in the collisions are given by $\sqrt{2GM_{wd}/R_{wd}}\sim 4\times 10^3$km/s,
where $M_{wd}=0.5M_{\odot}$ and $R_{wd}=10^4$km denote the typical mass 
and radius of the white dwarfs, respectively.
Furthermore, the bandwidths of the radiations are wider than those of radiations
emitted by neutron stars. This is 
because the magnetic fields of the white dwarfs are much weaker than those of neutron stars.
Thus, we can distinguish them from the radiations emitted from neutron stars.
The production rate of the bursts is
larger than the one of the FRBs observed,
if the number of the white dwarfs with strong magnetic fields $B\ge 10^9$G
is larger than $10^7$ in a galaxy. 
Furthermore, their luminocities are much larger than those of the FRBs observed. 
On the other hand,
the number of typical white dwarfs with magnetic fields $\sim 10^7$G is
much larger than the one of the white dwarfs with $B\ge 10^9$G.
However, it is difficult to observe the radiations from the white dwarfs with small magnetic fields $\le 10^7$G
because the total amount of the radiation energies is 
is not sufficientlly large to be observable.

In the next section (\ref{2}), we derive approximate solutions of axion stars with small masses
coupled with gravity.
We can see that the axion field oscillates with the frequency $m_a/2\pi$.
In the section (\ref{3}), we show that electric fields are produced on axion stars
under external magnetic fields. They are parallel to the magnetic fields 
and oscillate with frequency $m_a/2\pi$. 
In the section (\ref{4}), we determine masses of axion stars by comparison of
theoretical with observed rates of FRBs. The masses are found to be coincident with 
those estimated previously as the masses of the axion miniclusters.
In the section (\ref{5}), we describe how the radiations are emitted from the collisions.
Especially, we show that they are emitted from 
atmospheres of neutron stars. 
We find that once the axion stars touch the atmospheres,
they lose their energies by emitting radiations. 
In the section (\ref{6}), we show by calculating optical depth 
that the atmospheres are transparent for the radiations.
We also discuss that the radiations are circularly polarized after
they pass magnetospheres of neutron stars.
In the section (\ref{7}), we discuss how the thermal fluctuations of electrons emitting FRBs
give rise to narrow bandwidths of the radiations.
In the section (\ref{8}), we discuss that similar RFBs arise in the collisions between axion stars
and magnetic white dwarfs. 
We find that their durations are of the order of $0.1$ second
and their frequencies have bandwidth wider than those of FRBs from neutron stars.
We summarize our results in the final section (\ref{9}).

\section{axion stars}
\label{2}
First we would like to make a brief review of axions and axion stars.
Axions described by a real scalar field $a$ are Nambu-Goldstone boson 
associated with Pecci-Quinn global U(1) symmetry\cite{PQ}.
The symmetry was introduced to cure strong CP problems in QCD.
After the breakdown of the Pecci-Quinn symmetry at the period of much higher temperature than $1$GeV, 
axions are thermally produced as massless particles in the early Universe.
They are however only minor components of dark matter.
Since the axions interact with instanton density in QCD,
the potential term $-f_a^2m_a^2\cos(a/f_a)$
develops owing to instanton effects at the temperature below $1$GeV;
$f_a$ denotes the decay constant of the axions. 
Thus, the axion field oscillates around the minimum $a=0$ of the potential.
But, the initial value of the field $a$ at the temperature $1$GeV is unknown. It can take
a different value in a region from those in other regions causally disconnected.  
Thus, there are many regions causally disconnected 
at the epoch around the temperature $1$GeV, 
in each of which the axion field takes
a different initial value; energy density is also
different.
With the expansion of the Universe, the regions with different energy densities
are causally connected. Thus, there arises spatial fluctuations of the axion energy density.
The nonlinear effects of the axion potential cause the fluctuations with
over densities in some regions
grow to form axion miniclusters\cite{kolb} at
the period of equal axion-matter radiation energy density in the regions.
Their masses have been estimated to be of the order of $10^{-12}M_{\odot}$.
Furthermore, these miniclusters condense to form axion stars
with gravitationally losing their kinetic energies\cite{axions}.
Therefore, masses of axion stars are expected to be of the order of $10^{-12}M_{\odot}$.

\vspace{0.1cm}
Now we explain the classical solutions of the axion stars obtained 
in previous papers\cite{axions,osc,iwa,osci}.
The solutions are found by solving classical equations of axion field $a(\vec{x},t)$ coupled with gravity.
In particular we would like to obtain spherical symmetric 
solutions of axion stars with much smaller masses
than the critical mass\cite{osci} 
$M_{\rm max}$ of the axion stars; axion stars are stable when their masses are smaller than the 
critical mass $M_{\rm max}\sim 0.6\,m_{\rm pl}^2/m_a\simeq 0.5\times 10^{-5}M_{\odot}(10^{-5}\mbox{eV}/m_a)$.
The mass is obtained only for free axion field without self-interaction.
The gravity of the axion stars with much small masses
is much weak so that we may take the space-time metrics given by 

\begin{equation}
ds^2=(1+h_t)dt^2-(1+h_r)dr^2-r^2(d\theta^2+\sin\theta d\phi^2)
\end{equation} 
with both $h_t$ and $h_r \ll 1$.
It is easy to derive the equations of motion of the axion field and gravity,

\begin{eqnarray}
\label{eq}
(1-h_t)\partial_0^2 a&=&\frac{\partial_0 h_t-\partial_0 h_r}{2}\partial_0 a
+(1-h_r)(\partial_r^2+\frac{2}{r}\partial_r)a+\frac{\partial_r h_t-\partial_r h_r}{2}\partial_r a-m_a^2 a 
\nonumber \\
\partial_rh_t&=&\frac{h_r}{r}+4\pi Gr\Big((\partial_r a)^2-m_a^2a^2+(\partial_0 a)^2\Big) \\
\partial_rh_r&=&-\frac{h_r}{r}+4\pi Gr\Big((\partial_r a)^2+m_a^2a^2+(\partial_0 a)^2\Big) \nonumber
\end{eqnarray}
with the gravitational constant $G$,
where we assume the axion potential such that $V_a=-f_a^2m_a^2\cos(a/f_a)\simeq -f_a^2m_a^2+m_a^2a^2/2$
for $a/f_a\ll 1$.
As we will see later, the assumption holds for the axion stars with small masses, e.g. $10^{-12}M_{\odot}$. 

It is well known that there are no static solutions of real scalar fields coupled with gravity.
The fields and the metrics oscillate with time.
We expand the axion field and the metric such that

\begin{eqnarray}
a(t,r)&=&\sum_{n=0,1,2,,,}a_n(r)\cos((2n+1)\omega t)=a(r)\cos(\omega t)+a_1(r)\cos(3\omega t), \, ,\, ,\nonumber \\
h_{t,r}(t,r)&=&\sum_{n=0,1,2,,,}h^n_{t,r}(r)\cos(2n\omega t)=h_{t,r}^0(r)+h_{t,r}^1(r)\cos(2\omega t), \, ,\, ,
\end{eqnarray}
with $a(r)\equiv a_0(r)$. 
Then,
only by taking the terms proportional to $\cos^0(\omega t)=1$ and $\cos(\omega t)$ 
in the equations (\ref{eq}), 
we obtain

\begin{eqnarray}
-\omega^2(1-h^0_t+\frac{h^1_r-h^1_t}{2})a(r)&=&-\omega^2\frac{h^1_r-h^1_t}{2} a(r)
+(1-h^0_r)(\partial_r^2+\frac{2\partial_r}{r})a(r)-m_a^2a(r) \\ \nonumber
&& \quad \quad \quad \quad 
-\frac{1}{2}(\partial_rh^0_r-\partial_rh^0_t+\frac{\partial_rh^1_r-\partial_rh^1_t}{2})\partial_r a(r)\\
\partial_rh^0_t&=&\frac{h^0_r}{r}+2\pi Gr\Big((\partial_r a(r))^2-m_a^2a^2(r)+\omega^2 a^2(r)\Big)\\
\partial_rh^0_r&=&-\frac{h^0_r}{r}+2\pi Gr\Big((\partial_r a(r))^2+m_a^2a^2(r)+\omega^2 a^2(r)\Big)\\
\partial_rh^1_t&=&\frac{h^1_r}{r}+2\pi Gr \Big((\partial_r a(r))^2-m_a^2a^2(r)-\omega^2 a^2(r)\Big)\\
\partial_rh^1_r&=&-\frac{h^1_r}{r}+2\pi Gr\Big((\partial_r a(r))^2+m_a^2a^2(r)-\omega^2 a^2(r)\Big).
\end{eqnarray}

We note that when the gravitational effects vanish i.e. $G \to 0$, 
there is a solution $a=\tilde{a}_0\cos(m_at)$ 
with $\omega=m_a$ as well as $h_{t,r}^{0,1}=0$.
Since we consider axion stars with small masses, 
$\omega$ is almost equal to $m_a$; $m_a^2-\omega^2 \ll m_a^2$. That is, 
the binding energies $m_a-\omega$ are much smaller than $m_a$.
Furthermore, as we will see later, 
the radius $R_a$ of the axion stars with small masses is very large; $R_a \gg m_a^{-1}$.  
Thus, the term $\partial_ra(r)$ is much smaller than the term $m_a a(r)$,
i.e. $(\partial_ra(r))^2\ll (m_aa(r))^2$.
We may approximate the above equations in the following,

\begin{eqnarray}
\label{11}
(m_a^2-\omega^2)a(r)+m_a^2h^0_ta(r)&=&(\partial_r^2+\frac{2\partial_r}{r})a(r) \\
\partial_rh^0_t&\simeq &\frac{h^0_r}{r}\\
\partial_rh^0_r&\simeq &-\frac{h^0_r}{r}+4\pi Gr m_a^2a^2(r) \\
\partial_rh^1_t&\simeq &\frac{h^1_r}{r}-4\pi Gr m_a^2a^2(r) \\
\end{eqnarray}
with $h^1_r \ll h^0_{t,r} \ll 1$ and $h^1_r \ll h^1_t \ll 1$,
since $h^1_r \sim O\big(Gr^2((\partial_ra)^2+(m_a^2-\omega^2)a^2\big)$ and the other 
metrics $h \sim O\big(Gr^2m_a^2a^2\big)$. 

Therefore, we obtain the equation of the axion field,

\begin{equation}
\label{a}
-\frac{k^2}{2m_a}a(r)=-\frac{1}{2m_a}(\partial_r^2+\frac{2\partial_r}{r})a(r)+m_a\phi a(r)
\end{equation}
with $k^2\equiv m_a^2-\omega^2$,
where ``gravitational potential" $\phi\equiv h^0_t/2$ satisfies 

\begin{equation}
\label{gp}
(\partial_r^2+\frac{2\partial_r}{r})\phi =2\pi Gm_a^2a^2(r).
\end{equation}

Obviously, $k^2/2m_a$ represents a binding energy of the axion bounded to
an axion star whose mass $M_a$ is given by 
$M_a=\int d^3x ((\partial_0 a)^2+(\partial_r a)^2+m_a^2a^2)/2\simeq \int d^3x m_a^2a(t,r)^2
=\int d^3x m_a^2a(r)^2/2$ with the average taken in time;
$\omega^2\simeq m_a^2$.
The equation (\ref{a}) can be rewritten in the limit $r\to \infty $ as

\begin{equation}
\label{a1}
-\frac{k^2}{2m_a}a(r)=-\frac{1}{2m_a}(\partial_r^2+\frac{2\partial_r}{r})a(r)-\frac{Gm_aM_a}{r}a(r).
\end{equation}
A solution in eq(\ref{a1}) is given by $a(r)=\tilde{a}_0\exp(-kr)$ with $k=Gm_a^2M_a$.
Thus, we find that the radius $R_a=k^{-1}=(Gm_a^2M_a)^{-1}$ of the axion star is much larger than
$m_a^{-1}$ for small mass $M_a$. We can confirm numerically that the solution in eq(\ref{a1}) represents
approximate solutions of the equations (\ref{a}) and (\ref{gp}).
In this way 
we approximately obtain spherical symmetric solutions,

\begin{equation}
\label{a2}
a(\vec{x},t)=a_0f_a\exp(-\frac{r}{R_a})\cos(m_at),
\end{equation} 
with $r=|\vec{x}|$.
The solutions represent boson stars made of the axions bounded gravitationally, named as axion stars.
The solutions are valid for the axion stars with small masses $M_a\ll 10^{-5}M_{\odot}$.
The radius $R_a$ of the axion stars is numerically given in terms of the mass $M_a$ by

\begin{equation}
\label{R}
R_a=\frac{m^2_{\rm pl}}{m_a^2M_a}
\simeq 260\,\mbox{km}\,\,\Big(\frac{10^{-5}\rm eV}{m_a}\Big)^2\Big(\frac{10^{-12}M_{\odot}}{M_a}\Big), 
\end{equation} 
with the Planck mass $m_{\rm pl}$.  
The coefficient $a_0$ can be obtained by using 
the relations $M_a\simeq \int d^3x m_a^2a(r)^2/2=\pi m_a^2a_0^2f_a^2R_a^3/4$
and $m_a\simeq 6\times 10^{-6}\mbox{eV}\times (10^{12}\mbox{GeV}/f_a)$,
where the average is taken in time,  

\begin{equation}
\label{a_0}
a_0\simeq 0.9\times 10^{-6} \Big(\frac{10^2\,\mbox{km}}{R_a}\Big)^2\frac{10^{-5}\mbox{eV}}{m_a}.
\end{equation}
Thus, the condition $a/f_a\ll 1$ is satisfied 
for the axion stars with small mass $M_a \sim 10^{-12}M_{\odot}$.
We have simply used the mass $10^{-12}M_{\odot}$ for reference. 
But, the mass is the one we determine by the comparison of the theoretical
and observational event rates of FRBs, as we show below.
The mass is much smaller than the critical mass $M_{\rm max}\sim 10^{-5}M_{\odot}$. 
Thus, the solutions represent stable axion stars.
In this way the solutions along with the parameters $R_a$ and $a_0$ can be approximately obtained. 
Obviously, the axion stars are composed of axions with much small momenta $\sim 1/R_a$.

\vspace{0.1cm}
Here we should make a comment on the quartic term $-(m_a^2/f_a^2)a^4/24$ of the potential 
$V_a=-f_a^2m_a^2\cos(a/f_a)\simeq -f_a^2m_a^2+m_a^2a^2/2-(m_a^2/f_a^2)a^4/24$.
We have neglected the term in the above discussion.
Although the term is much smaller than the mass term $m_a^2a^2/2$,
the term gives a contribution $-m_a^2a(r)^3/(12f_a^2)$ in the eq(\ref{11}) which is comparable to
the term $(\omega^2-m_a^2)a(r)\simeq R_a^{-2}a(r)$
when $R_a\simeq 130$km or less.
Since the quartic term is negative in the potential, the masses of the axion stars decrease
with the increase of the field amplitude $a_0$. That is, the axion stars become unstable
when the term is comparable to the term $(\omega^2-m_a^2)a(r)$. 
Thus, our solutions are only stable for the axion stars 
with larger radii than $130$km ( smaller masses than $0.2\times 10^{-11}M_{\odot}$.)
This indicates that the critical mass becomes much smaller than $M_{\rm max}$ 
when the quartic term is taken into account. We do not yet know real critical mass when 
we take account of the full potential $V_a$ of the axions.
But the axion stars at least with masses smaller than $0.2\times 10^{-11}M_{\odot}$
are stable since the approximation of neglecting the quartic term is valid.
( Much small critical masses $\sim 10^{-21}M_{\odot}$ have been 
previously pointed out\cite{barranco}
using the procedure in the previous work\cite{ruffini}.
But the procedure
is only valid for free fields coupled with gravity.
Especially, it is not applicable for the real scalar fields with nonlinear interactions
such as the axions. 
The approximations such as $<\hat{a}^4>=<\hat{a}^2><\hat{a}^2>$ is used in the reference 
for obtaining the axion stars
where $\hat{a}$ represents axion field operator and the state $|>$ does a state 
with number of axions fixed, i.e. eigenstates of the axion number operator.
On the other hand, our classical approximation is to assume that the state $|>$ represents
a coherent state of axions. Thus, we have $<\hat{a}^4>=<\hat{a}>^4$.
The critical mass shown in the previous paper is of the order of $10^{-20}M_{\odot}$, while
as we have shown perturbatively, the critical masses are 
of the order of $10^{-12}M_{\odot}$. The difference comes from the use of the different approximations.
We point out that the critical masses we obtain are of the same order of the magnitude
as the ones shown in the paper\cite{t}. 
Anyway, more rigorous treatments are needed to see the precise values of the critical mass. ) 


In our analysis,
the radius $130$km of the axion stars with the mass
$0.2\times 10^{-11}M_{\odot}$ was roughly derived only as a guide
for a critical radius.  
We may use the radius $10^2$km as a reference
of the stable axion stars in the discussion below.

\section{axion stars in magnetic fields}
\label{3}
We proceed to discuss electric field $\vec{E}_a$ generated on the axion stars 
under magnetic field $\vec{B}$. It is well-known that
the axion couples with both electric $\vec{E}$ and magnetic fields $\vec{B}$ in the following,

\begin{equation}
\label{L}
L_{aEB}=k\alpha \frac{a(\vec{x},t)\vec{E}\cdot\vec{B}}{f_a\pi}+\frac{\vec{E}^2-\vec{B}^2}{2}
\end{equation}
with the fine structure constant $\alpha\simeq 1/137$,   
where the numerical constant $k$ depends on axion models; typically it is of the order of one.
Hereafter we set $k=1$.
From the Lagrangian, we derive the Gauss law, 
$\vec{\partial}\cdot \vec{E}=-\alpha\vec{\partial}(a\vec{B})/f_a\pi$. Thus,
the electric field generated on the axion stars under the magnetic field $\vec{B}$
is given by 

\begin{eqnarray}
\label{ele}
\vec{E}_a(r,t)&=&-\alpha \frac{a(\vec{x},t)\vec{B}}{f_a\pi}
=-\alpha \frac{a_0\exp(-r/R_a)\cos(m_at)\vec{B}(\vec{r})}{\pi}\\
&\simeq& 0.4\times 1\mbox{eV}^2( \,\,=2\times 10^4\mbox{eV}/\mbox{cm} \,\,)\cos(m_at)
\Big(\frac{10^2\,\mbox{km}}{R_a}\Big)^2\frac{10^{-5}\mbox{eV}}{m_a}\frac{B}{10^{10}\mbox{G}}.
\end{eqnarray} 
 We find that the electric field is very strong at neutron stars with magnetic fields $\sim 10^{10}$G,
while it is much weak at the sun with magnetic field $\sim 1$G. 
The electric field $\vec{E_a}$ is parallel to the magnetic field $\vec{B}$ and
oscillates coherently over the whole of the axion stars. 
When the axion stars are in magnetized ionized gases,
the field induces coherently oscillating electric currents with large length scale $R_a$ 
of the axion stars. Thus the large amount of dipole radiations can be emitted. 
Especially, electrons in the atmospheres of neutron stars oscillate 
and emit coherent dipole radiations with the frequency $m_a/2\pi$. 
We should mention that the motions of charged particles accelerated by the electric field $\vec{E_a}$ 
are not affected by the magnetic field $\vec{B}$ since $\vec{B}$
is parallel to $\vec{E}$.
Thus, they can emit dipole radiations.  

Here we make a comment that the electric fields in eq(\ref{ele}) are the ones generated 
at the rest frame of the axion stars. When the axion stars collide with neutron stars,
the magnetic field $\vec{b}=\vec{v}_c\times \vec{E}_a$ 
with $v_c^2\ll 1$ is induced at the rest frame of the neutron stars,
where $\vec{v}_c$ represents a relative velocity between the axion stars and the neutron stars.
Since $v_c\sim 0.1$, the magnetic field $\vec{b}$ is much smaller than $\vec{B}$.
Thus, the effect can be neglected.
Similarly,
we can neglect the effects of the rotations of the neutron stars,
whose velocities 
are much smaller than 
the relative velocities $v_c$.

As will be shown later,
all the energy of a part of the axion star 
touching the atmospheres of neutron stars  
is released into the radiations, since the atmospheres of neutron stars
are composed of highly dense electrons and ions. The electric fields of the axion stars
induce oscillating electric currents which produce the radiations. 
Namely, the neutron stars make the axion stars evaporate into the radiations. 
The frequency of the radiations is given by $m_a/2\pi\simeq 2.4\times(10^{-5}\mbox{eV}/m_a)$GHz
at the rest frame of the axion stars.
Therefore, when the axion stars collide with neutron stars,
the large amount of the radiations is produced within 
a short period $R_a/v_c$ being of the order of milli seconds; $v_c$ is of the order of $10^5$km/s, see later. 
These radiations can escape the atmospheres and magnetosphere of neutron stars,
because they are optically thin for the radiations as we show below. 
Thus, it is reasonable to identify the radiations as the FRBs observed.

\section{event rate of fast radio bursts}
\label{4}
We calculate the rate of the collisions between axion stars and neutron stars in a galaxy.
The collisions generate FRBs so that the rate is the event rate of the FRBs.
By the comparison of theoretical
with observed rate of the bursts, we can determine the mass of the axion stars.
We assume that halo of a galaxy is composed of the axion stars whose
velocities $v$ relative to neutron stars is supposed to be $3\times 10^2 $\,km/s. 
Since the local density of the halo is supposed to be $0.5\times 10^{-24}\,\mbox{g\,cm}^{-3}$,
the number density $n_a$ of the axion stars is given by $n_a=0.5\times 10^{-24}\,\mbox{g\,cm}^{-3}/M_a$. 
The event rate $R_{\rm burst}$ can be obtained in the following,

\begin{equation}
R_{\rm burst}=n_a\times N_{\rm ns}\times Sv\times 1\rm year,
\end{equation}
where $N_{\rm ns}$ represents the number of neutron stars in a galaxy;
it is supposed to be $10^9$.
The cross section $S$ for the collision is given by
$S=\pi (R_a+R_{ns})^2\big(1+2GM_{ns}/v^2(R_a+R_{ns})\big)\simeq 2.8\pi (R_a+R_{ns})GM_{\odot}/v^2$ 
where $R_{ns}\,(=10$km ) denotes the radius of neutron star
with mass $M_{ns}=1.4M_{\odot}$. 
It follows that the observed event rate is given by

\begin{eqnarray}
R_{\rm burst}&=&\frac{0.5\times 10^{-24}\,\mbox{g\,cm}^{-3}}{M_a}\times 10^9\times 
2.8\pi(10\mbox{km}+R_a)\frac{GM_{\odot}}{10^{-6}}\times 1\rm year \nonumber \\ 
&\sim& 
10^{-3}\Big(\frac{10^{-12}M_{\odot}}{M_a}\Big)\frac{
10\mbox{km}+260\mbox{km}\Big(\frac{10^{-5}\mbox{eV}}{m_a}\Big)^2\frac{10^{-12}M_{\odot}}{M_a}}
{10\mbox{km}+260\mbox{km}}.
\label{Rb}
\end{eqnarray}
Therefore, we can determine the masses $M_a$ of the axion stars by the comparison of $R_{\rm burst}$ in eq(\ref{Rb})
with
the observed event rate $\sim 10^{-3}$ per year in a galaxy. We obtain
$M_a\sim 10^{-12}M_{\odot}$ when $m_a=10^{-5}$eV.
The parameters used above still involves large ambiguities. Furthermore, 
the rate becomes larger than
that in eq(\ref{Rb}) when we take into account the cosmological evolution of the Universe. Thus,
the observed rate only constrains the masses of the axion stars in a range such that 
$M_a=10^{-12}M_{\odot}\sim 10^{-11}M_{\odot}$.
It is remarkable that
the mass $M_a\sim 10^{-12}M_{\odot}$ of the axion stars obtained  
is coincident
with the masses of axion miniclusters\cite{mini} estimated previously.

Using the formula eq(\ref{R}) we find  
the radius $\sim 10^2$\,km of the axion stars with the mass $\sim 10^{-12}M_{\odot}$,
which is larger than those of
neutron stars.
Then, when the collisions take place, the neutron stars pass through the insides of the axion stars.
As we will show later, when the axion stars touch the atmospheres of the neutron stars,
the large amount of the radiations is emitted instantaneously so that the parts of
the stars touching them lose their energies. 
The amount of the radiation energy released in the collision is given by
$10^{-12}M_{\odot}(10\rm km/10^2\rm km)^2\sim 10^{43}$GeV.
Thus, our production mechanism of FRBs can explain  
the observed energies of the FRBs.

\section{radiations from axion stars in atmospheres of neutron stars}
\label{5}
We estimate how large amount of energies the axion stars emit as radiations in the collisions
with neutron stars. In particular, 
we show that they rapidly lose their energies in the atmospheres\cite{atmos} of neutron stars.
We consider old neutron stars which are dominant components of neutron stars in
the Universe. Their temperatures ( magnetic fields ) are assumed
to be of the order of $10^{5}$K ( $10^{10}$G ). We also assume that the neutron stars have
hydrogen atmospheres.

First, we show how an electron emit radiations in the electric fields of the axion stars.
The electric field $\vec{E}_a$ on the axion stars 
generated under magnetic fields makes an electron oscillate according to the
equation of motion $\vec{\dot{p}}=(-e)\vec{E}_a+(-e)\vec{v}\times \vec{B}+m_e\vec{g}$ 
with electron mass $m_e$, 
where $\vec{p}$ and $\vec{v}$
denote momentum and velocity of the electron, respectively and $\vec{g}$ does
surface gravity of neutron stars; $|\vec{g}|=M_{ns}G/R_{ns}^2$. 
We note that the electric field $\vec{E}_a$ 
is parallel to the magnetic field $\vec{B}$. Thus,
the direction of the oscillation is parallel to $\vec{B}$. The magnetic field does not affect
the oscillation.
Similarly, the gravitational forces does not affect it since they are much weaker than the electric fields. 

Then, the equation of motion of the electron parallel to $\vec{E}_a$ is 
given by $\dot{p}=-eE_a$.
Since the electric fields oscillate such as $E_a\propto \cos(m_at)$,
the electron oscillates with the frequency $m_a/2\pi$ so that it emits a dipole radiation.
The amplitude of the oscillator is given by $e\alpha a_0B/(m_a^2 m_e\pi)\simeq 0.05$cm
which is smaller than the wave length 
$\lambda\sim 10\mbox{cm}(10^{-5}\mbox{eV}/m_a)$ of the radiations.
Thus, 
the emission rate of the radiation energy produced by a single electron with the mass $m_e$ is
given by 

\begin{equation}
\dot{w}\equiv \frac{2e^2\dot{p}^2}{3 m_e^2}=\frac{2e^2(e\alpha a_0 B/\pi)^2}{3m_e^2}
\simeq 0.7\times 10^{-9}\mbox{GeV/s}\Big(\frac{10^2\rm km}{R_a}\Big)^4\Big(\frac{10^{-5}\rm eV}{m_a}\Big)^2
\Big(\frac{B}{10^{10}\mbox{G}}\Big)^2.
\end{equation}

Electrons coherently oscillate in the volume $\lambda^3$, in which
there exist a number of the electrons with their number $N_e=n_e\lambda^3$
where $n_e$ denotes the number density of electrons.
Then, the total emission rate $\dot{W}$ from the electron gas is given such that
 $\dot{W}=\dot{w}(n_e\lambda^3)^2=2(n_e\lambda^3)^2\dot{p}^2/(3 m_e^2)$. 
On the other hand, if the depth $d$ of the atmosphere of neutron stars 
is less than the wave length of the radiations,
the number of electrons coherently oscillating is given by $n_ed\lambda^2$. 
Actually, the depth $d$ of the hydrogen atmosphere with temperature of the order of $10^5$K
is about $0.1$cm, which is much smaller than the wave length
$\sim 10\mbox{cm}(10^{-5}\mbox{eV}/m_a)$. 

We make a comment that the thermal effects of electron gas under consideration 
do not disturbe the oscillation by the electric field. 
Since the temperatures of the atmospheres are 
supposed to be $10^5$K, the thermal energy $\sim 10$eV of an electron is 
much smaller than the kinetic energy of the oscillation, 
$p^2/2m_e=(eE)^2/2m_em_a^2\sim 10^2\mbox{eV}(B/10^{10}\mbox{G})^2$ with $m_a=10^{-5}$eV.
Although the oscillation is never disturbed in the thermal bath,
the frequency of the radiations recieves the effect of the thermal fluctuations
so that the radiations have finite bandwidth. 

We also make a comment about the depth $d$ of atmospheres of neutron stars.
We suppose that the density distribution $\rho(r)$ is given by
$\rho(r)=\rho_0\exp(-r/d)$ 
with the depth $d=k_BT/mg$ 
( $m$ denotes average mass of the atoms composing the atmospheres,
$T$ does temperature of the atmospheres and $k_B$ does Boltzmann constant. ) 
The distribution may be obtained by
solving the equation of the dynamical valance 
$\partial_rP(r)=-\rho(r) g$ between pressure $P$ and surface gravity $g\equiv GM/R^2$ 
with the use of the equation of state $P(r)=n(r)k_BT$ of ideal gas. 
Here $T$ denotes the constant temperature and 
$M$ ( $R$ ) and $n$ denote mass ( radius ) of the star and 
number density ( $n=\rho/m$ ) of atoms composing the atmosphere.
For example, $d\sim 10$km for $T=300$K, $g_e=9.8$m/s$^2$ and $m=28$GeV in the case of the earth,
while $d\sim 0.1$cm for 
$T=10^5$K, $g_n=10^{11}\times g_e$ and $m=1$GeV in the case of hydrogen atmosphere of neutron stars.   
Although the estimation is very rough, we can grip on
the depth of the atmosphere of the neutron stars, which is given by $0.1$cm when the temperature
is of the order of $10^5$K. We can see that the number density of electrons 
$n_e(r)=n_0\exp(-r/0.1\mbox{cm})$
decreases rapidly with the distance $r$ from the bottom of the atmospheres.
The radiations emitted in the atmospheres can pass through the atmospheres
without absorption because the atmosphere is transparent for the radiations
with transverse polarizations, as shown in the next section.

\vspace{0.1cm}
We proceed to show that the axion stars rapidly evaporate into the radiations 
when they touch the atmospheres of neutron stars.
We assume that the atmospheres are composed of fully ionized hydrogen gas 
with temperature of the order of $10^5$K,
whose depth $d$ is about $\sim 0.1$cm. Thus, we have
the density distribution $n_e(r)=n_0\exp(-r/0.1\mbox{cm})$
where the density $n_e(r=0)=n_0$ at the bottom is much larger than $10^{24}$/cm$^3$. 
In the paper we take $n_0=10^{24}$/cm$^3$.
It approximately corresponds to
the density $1$g/cm$^3 $.  
We consider the radiations arising from a region 
with volume $d\lambda^2\sim 10\mbox{cm}^3$ in the atmospheres.
The emission rate $\dot{W}$ of the radiations from the region is given by,

\begin{equation}
\dot{W}\sim 10^{-9}(d\lambda^2 n_e)^2\mbox{GeV/s}\,\Big(\frac{B}{10^{10}\mbox{G}}\Big)^2 
\sim 10^{37}\,\mbox{GeV/s}\,\Big(\frac{n_e}{10^{22}\mbox{cm}^{-3}}\Big)^2
\,\Big(\frac{10^2\rm km}{R_a}\Big)^4\Big(\frac{10^{-5}\rm eV}{m_a}\Big)^6\Big(\frac{B}{10^{10}\mbox{G}}\Big)^2,
\end{equation}
where we have taken, for instance, the number density $n_e=10^{22}/\rm cm^3$ of electrons 
in the region with the density $\rho\sim 10^{-2}$g/cm$^3$, 
which is located roughly at the height $r=0.5$cm. ( As we take larger $n_e$, $\dot{W}$ becomes
larger. )
On the other hand, the energy of the axion stars contained in the volume $d\lambda^2=10$cm$^3$
is given by $10^{-12}M_{\odot}10\mbox{cm}^3/(4\pi R_a^3/3)\sim 10^{24}\mbox{GeV}$.
This energy is smaller than the energy of the radiations
$\dot{W}\times 10^{-11}\rm s\simeq 10^{26}$GeV emitted within
a time $0.1\mbox{cm}/v_e \sim 10^{-11}$s in which the axion stars pass the depth $d=0.1$cm. 
It should be noted that
the relative velocity $v_e$ of the axion stars when they collide with the neutron stars, 
is given by $v_e=\sqrt{2G(1.4M_{\odot})/R_{ns}}\simeq 6\times 10^{-1}\simeq 2\times 10^5$km/s.
Therefore, we find that the whole energy of the region 
with the volume $\lambda^2 d$ in the axion stars is transformed into 
the radiation energy when the region pass through the atmospheres.

The purpose using the specific values $n_e=10^{22}$cm$^{-3}$ or the depth $d=0.1$cm 
in the estimation is simply to show that 
the whole energies of the part of the axion stars passed through by neutron stars
are transformed into radiations. 
Obviously, our results do not depend on the specific values.
The use of different values similar to these ones does not change our results.  
Therefore, we conclude that the part of
the axion stars touching the neutron stars
are immediately evaporated into the radiations.

There are ambiguities about the parameters ( temperature, density, composition, e.t.c ) 
of neutron star atmospheres. Only what we need to derive our results is the fact that
the average number density of electrons is much large such as
$10^{22}/\mbox{cm}^3$
in the atmosphere and that strong magnetic fields $\ge 10^{10}$G is present.
These assumptions
are generally acceptable.
Thus, our production mechanism of
the FRBs is fairly promising. 
 
\section{transparency of neutron star atmosphere} 
\label{6}
The radiations produced in the atmospheres 
can pass through them and arrive at the earth. They are never absorbed within the
atmospheres.
We will show that the atmospheres are transparent 
for the radiations, even if $n_e\simeq 10^{22}\mbox{cm}^{-3}$ when 
the strong magnetic field stronger than $B=10^{10}$G is present.
We assume that the temperatures of the atmospheres are of the order of $10^5$K and
that electron density is given by $n_e(r)=n_0\exp(-r/0.1\rm cm)$,
where the depth of the atmospheres is taken as $0.1$cm, as was
shown above.
We also assume for simplicity 
that the atmospheres are composed of fully ionized hydrogen atoms.

Then, we may use the following formula\cite{abs} of the free-free absorption coefficient,

\begin{equation}
C_{\epsilon}(r)=\frac{n_e(r)}{(\omega+\epsilon\,\omega_c)^2+
\nu_{\epsilon}(r)^2}\frac{4\pi e^2\nu_{\epsilon}(r)}{m_e}
\end{equation}
with $\omega=m_a/2\pi$, $\omega_c=eB/m_e$ and $\omega_p=eB/m_p$,
where $m_p$ denotes proton mass and $\nu_{\epsilon}$ is given by

\begin{equation}
\nu_{\epsilon}(r)=\frac{2e^2\omega^2}{3m_e}+
\frac{4n_e(r) e^4\Lambda_{\epsilon}(T,B,\omega )}{3T}\sqrt{\frac{2\pi}{m_eT}}
\end{equation} 
where we used $\omega/T \ll 1$ since $T=10^5$K and $m_a=10^{-5}$eV. 
The parameter $\epsilon=0,\pm $ denotes three types of polarizations;
circular polarizations $\epsilon=\pm $ ( polarized transverse to $\vec{B}$ ) 
and longitudinal polarization $\epsilon=0$ ( polarized longitudinal to $\vec{B}$ ).
The explicit formula of 
$\Lambda_{\epsilon}(T,B,\omega )$ is given by

\begin{equation}
\Lambda_{\epsilon}(T,B,\omega )=
\frac{3}{4}\sum_{n=-\infty}^{\infty}\int_{0}^{\infty}Q_{\epsilon}(n,T,B,\omega, y)dy
\end{equation}

where

\begin{eqnarray}
Q_{\epsilon}(n,T,B,\omega,y)&=&\frac{yA_n^{\epsilon}(T,B,\omega, y)}
{\sqrt{1+2\theta y+y^2}\bigl(y+\theta+\sqrt{1+2\theta y+y^2}\bigr)^{|n|}\bigl(\sinh(b/2)\bigr)^{|n|}}\\
A_n^0(T,B,\omega, y)&=&\frac{x_nK_1(x_n)}{y+b/4},\quad 
A_n^{\pm}(T,B,\omega, y)=\frac{\omega^2}{(\omega\mp \omega_p)^2}\frac{(y+\theta+|n|\sqrt{1+2\theta y+y^2})K_0(x_n)}{1+2\theta y+y^2}\\
b&=&13.6\frac{B}{10^{10}\mbox{G}}\frac{10^5\rm K}{T},
\quad x_n=|\omega/T-nb|\sqrt{0.25+y/b},
\quad \theta =\frac{1+\exp(-b)}{1+\exp(-b)}\simeq 1
\end{eqnarray} 
with $x_0\simeq |\omega /T|\sqrt{0.25+y/b}$ and $x_{n\neq 0}=|nb|\sqrt{0.25+y/b}$
since $\omega/T\simeq 10^{-7}$.
$K_0$ and $K_1$ represent modified Bessel functions.

Here we note that the contributions of the sum over large integer $n$ are very small
because there are damping factors such as $1/(2^{|n|}\sinh(b/2)^{|n|})\simeq 1/(800)^{|n|}$ 
and integrands of $y$
have the factor $\exp(-|n|\sqrt{yb})$ for large $y$. 
The integration $\int^{\infty} dy \exp(-|n|\sqrt{yb})$ gives a damping factor $n^{-2}$ for
large $n$.
Thus, the main contribution comes from the integral of 
$\int_0^{\infty}Q_{\epsilon}(n=0,T,B,\omega,y)dy$. We should also note the presence of 
the small factor $\omega^2/(\omega\pm\omega_p)^2\simeq \omega^2/\omega_p^2\simeq 10^{-7}(10^{10}\rm G/B)^2$ in $A_n^{\pm}$.
The factor comes from the finiteness of proton mass; the recoil effect of the proton owing to
the absorption of the radiations.
Therefore, the absorption coefficient can be approximately rewritten by

\begin{equation}
C_{\pm}(r)\simeq \frac{n_e(r)}{\omega_c^2}\frac{4\pi e^2\nu_{\pm}(r)}{m_e}
\end{equation}
where

\begin{equation}
\nu_{\pm}(r)\simeq 
\frac{4n_e(r) e^4\Lambda_{\pm}(T,B,\omega )}{3T}\sqrt{\frac{2\pi}{m_eT}}\simeq
\frac{n_e(r) e^4\int_0^{\infty}Q_{\pm}(n=0,T,B,\omega)}{T}\sqrt{\frac{2\pi}{m_eT}}
\end{equation}
with $\theta\simeq 1$ and

\begin{equation}
Q_{\pm}(n=0,T,B,\omega)\simeq \frac{\omega^2}{\omega_p^2}\frac{yK_0(|\omega /T|\sqrt{0.25+y/b})}{(1+y)^2}.
\end{equation}

On the other hand, the absorption coefficient $C_0(r)$ for longitudinally polarized radiations is
given by 

\begin{equation}
C_0(r)\simeq \frac{n_e(r)}{\omega_c^2}\frac{4\pi e^2\nu_0(r)}{m_e}
\end{equation}
where 
 
\begin{equation}
\nu_0(r)\simeq 
\frac{4n_e(r) e^4\Lambda_0(T,B,\omega )}{3T}\sqrt{\frac{2\pi}{m_eT}}\simeq
\frac{n_e(r) e^4\int_0^{\infty}Q_0(n=0,T,B,\omega)}{T}\sqrt{\frac{2\pi}{m_eT}}
\end{equation}
with

\begin{equation}
Q_0(n=0,T,B,\omega)\simeq \frac{y|\omega /T|\sqrt{0.25+y/b}\,K_1(|\omega /T|\sqrt{0.25+y/b})}{(1+y)(y+b/4)}.
\end{equation}

Using these formulae,
we can see the optical depth $\tau_{\pm}(r_c)=\int_{r=r_c}^{\infty}dr'C_{\pm}(r')<1$ 
even at the location $r_c$ in which
the number density $n_e(r_c)$ is equal to $10^{22}$/cm$^3$.
Therefore, we find that
the atmospheres are transparent for the radiations with the circular polarizations.
The transparency comes from the fact that the frequency $\omega=m_a/2\pi$ is much smaller than 
the cyclotron frequencies $\omega_c$ and $\omega_p$ under the strong magnetic fields $B=10^{10}$G.
Physically, the electric fields of the radiations 
hardly make electrons move transversely to the direction of the magnetic fields $B$.
Thus, they cannot be absorbed.
On the other hand, we can easily see that 
the atmospheres are opaque ( $\int_{r=r_c}^{\infty}dr'C_0(r')\gg 1$ )
for the radiations with the longitudinal polarization
since the radiations easily make electrons oscillate longitudinally; they are absorbed
by the electrons.

\vspace{0.2cm}
We would like to mention
that although the radiations emitted from the atmospheres are linearly polarized,
some of them are circularly polarized after
they pass through the
magnetospheres of neutron stars.
The magnetospheres are composed of electrons or positrons, 
which are produced by the Schwinger mechanism under electric field
associated with the rotation of the magnetic field $B$.
The charged particles are distributed to screen the electric field. 
The number density of electrons ( positrons ) in the magnetospheres
is given by the Goldreich-Julian density 
$\simeq \Omega B/2\pi\sim 10^7$cm$^{-3}\big(\Omega/(2\pi/\mbox{s})\big)\big(B(r)/10^{10}\mbox{G}\big)$ 
with angular velocity $\Omega$ of
neutron stars.
These
electrons ( positrons ) absorb right ( left ) handed circularly polarized radiations
when the cyclotron frequency $\omega=eB(r)/m_e$ becomes equal to $m_a/2\pi$, respectively.
Since $B(r)$ decreases such that $B(r)\propto 1/r^3$,
the absorption arises around the location at the height $r_{ab}\sim 10^3$km above the surface of 
neutron stars. It implies that
the absorption coefficient $C_{\pm}(r_{ab})$ is much large for a type of circularly polarized radiations
compared with the one for the other type of circularly polarized radiations, 
for instance, $C_{+}(r_{ab}) \gg C_{-}(r_{ab})$.
The spatial distribution of the electrons is different from the distribution of the positrons.
Therefore, 
the radiations passing through the magnetospheres are
circularly polarized.
Such a polarization has been observed\cite{frb3} in FRB 140514.

\vspace{0.1cm}
We would like to point out that 
the atmospheres may evapolate instantaneously when the radiations pass them.
This is because
even if only a fraction of the radiation energies is dissipated in the atmospheres,
the energy is sufficiently large to make the atmospheres evapolate. For example,
a fraction e.g. $10^{-8}$ of the radiation energies $10^{43}$GeV,
( $10^{-8}\times 10^{43}\mbox{GeV}=10^{35}$GeV )
gives a large energy $10^{35}/(N=10^{35})=1$GeV to each nucleon in the atmospheres;
$N\sim 10^{24}/\rm cm^3 \times 0.1cm (10^6cm)^2=10^{35}$.
Then, it apparently seems that our production mechanism of the FRBs does not work.
But we should note that the FRBs are also produced in envelopes present
just below the atmospheres.
The envelopes are more dense ( $10^{24}\rm /cm^3\sim 10^{32}\rm /cm^3$ ) in electron number density 
and deeper ( $\sim 10^4$ cm ) than the atmospheres. Thus, even if the atmospheres instantaneously evapolate,
the radio bursts with sufficiently large energies as observed 
are produced in the envelopes of neutron stars.

\section{narrow bandwidth}
\label{7}
It apparently seems that the radiations are monochromatic, that is, their frequencies are
given by the axion mass. On the other hand, the FRBs have been observed with the frequencies 
in the range of $1.2$GHz$\sim 1.6$GHz. Here we would like to show that
the radiations emitted from the neutron stars have finite bandwidth 
including the range of the observed frequencies.
They are dipole radiations emitted by electrons harmonically oscillating.
These electrons have temperatures
of the order of $10^5\mbox{K}\simeq 10$ eV. Thus, 
we take account of 
thermal effects on the oscillations. The kinetic energies $\epsilon_k$ of the oscillations are 
given by $p^2/2m_e=(eE)^2/2m_em_a^2\sim 10^2\mbox{eV}(B/10^{10}\mbox{G})^2$ with $m_a=10^{-5}$eV. 
The energy is equal to the potential energy $m_e\omega^2x_e^2/2$ of the harmonic oscillations
with the frequency $\omega=m_a/2\pi$;
$x_e$ represents the amplitude of electrons.
When the thermal fluctuations are added to the harmonic oscillations,
the electron motion may be described by the following Langevin equation,

\begin{equation}
m_e\ddot{x_e}=-m_e\omega^2 x_e+\eta
\end{equation}
where the thermal fluctuation is represented by $\eta$.
Since we consider only the effect on the harmonic oscillation $x_e=x_0\cos(\omega t)$,
we take only the term of $\eta=\eta_0\cos(\omega't)$. Then, the frequency 
$\omega'=\omega+\omega_{\rm th}$ ( $\omega \gg \omega_{\rm th}$ ) of electrons
can be derived from the Langevin equation 
such that $m_e(\omega'^2-\omega^2)x_0\simeq 2m_e\omega\omega_{\rm th}x_0=\eta_0$;
$\omega_{\rm th}=\eta_0/(2m_e\omega x_0)$.
Thus, the fluctuations $\omega_{\rm th}$ in the frequency is obtained by taking average of  
the thermal fluctuation $\eta_0$ with an appropriate Gaussian distribution,

\begin{equation}
<\omega_{\rm th}^2>=\frac{1}{4m_e^2\omega^2x_0^2}<\eta_0^2>
=\frac{\omega^2<y^2>}{4x_0^2}=\frac{T}{2m_ex_0^2}=\frac{\omega^2 T}{4\epsilon_k},
\end{equation} 
with $\eta_0\equiv m_e\omega^2y$.
The Gaussian distribution is assumed to be given by $\exp(-m_e\omega^2y^2/4T)$.

 Thus, the thermal fluctuations in the frequencies of electrons are given by

\begin{equation}
\omega\pm \omega_{\rm th}=\omega(1\pm \frac{\omega_{\rm th}}{\omega})
=\frac{m_a}{2\pi}\Big(1\pm \sqrt{\frac{T}{4\epsilon_k}}\Big)\simeq 
\frac{m_a}{2\pi}\Big(1\pm \sqrt{\frac{10\rm eV}{4\times 10^2 \rm eV}}\Big)
\simeq \frac{m_a}{2\pi}(1\pm 0.16),
\end{equation}
where we take values $T=10^5$, $B=10^{10}$G and $m_a=10^{-5}$eV.
The thermal fluctuations of the electrons 
cause the finite but narrow bandwidth of the FRBs. 
We should note that the fluctuation $\omega_{\rm th}/\omega$ 
depends on the temperature $T$, magnetic field $B$
and axion mass $m_a$,

\begin{equation} 
\label{th}
\frac{\omega_{\rm th}}{\omega}=\sqrt{\frac{T}{4\epsilon_k}}
\sim 0.1\frac{m_a}{10^{-5}\rm eV}\frac{10^{10}G}{B}\sqrt{\frac{T}{4\rm eV}}.
\end{equation}

The equation is used for the estimation of the bandwidths of the radio bursts from
the collision between axion stars and white dwarfs, which is discussed in next section. 

\vspace{0.1cm}
We should mention that the observed radiations receive several redshifts.
The frequency of the electric fields induced on the axion stars under the magnetic fields
is equal to $\omega=m_a/2\pi$.
Since the axion stars collide with the neutron stars at the relative velocity $\sqrt{2GM_{ns}/R_{ns}}$,
the frequencies $\omega_{ns}$ of oscillating electrons induced by the electric fields
is given by
$\omega_{ns}=\omega\sqrt{1-2GM_{ns}/R_{ns}}$ at the rest frame of the neutron stars.
Thus, the radiations with the frequency $\omega_{ns}$
are emitted by the electrons at the rest frame of the neutron stars.
The radiations receive gravitational redshifts when we observe them
far from the neutron stars. The frequency $\omega'$ is
given by $\omega'=\omega_{ns}\sqrt{1-2GM_{ns}/R_{ns}}$.
Finally, the frequency of the radiations observed at the earth is
given by $\omega_{ob}=\omega'/(1+z)=\omega(1-2GM_{ns}/R_{ns})/(1+z)$
when the neutron stars are located at the places with redshift $z$.

\section{collisions with magnetic white dwarfs}
\label{8}
Up to now, we have considered that FRBs arise from the collisions between axion stars and neutron stars.
Similarly, FRBs may arise from the collisions between axion stars and magnetic white dwarfs\cite{wd}.
Some of the magnetic white dwarfs have strong magnetic fields such as $10^9$G.
They have dense hydrogen atmospheres with
temperatures of order of $10^4$K and depths of the order of $10^4$cm.
( We can easily estimate the depth by taking account of the physical parameters,
the surface gravity $g_{wd}\simeq 10^5\times g_e$ 
and the temperature $T=10^4$K of the white dwarfs. Thus, the density distribution is
given by $n=n_0\exp(-r/10^4\rm cm)$. )  
They have dense free electrons similar to the case of neutron stars.
Thus, by the collisions with axion stars, radiation bursts similar to the observed FRBs are produced.  
It turns out that the duration of the bursts is of the order of $0.1$ second. This is because
the axion stars collide with white dwarfs at the velocity $\sqrt{2GM_{wd}/R_{wd}}\simeq 4000$km/s
where the mass $M_{wd}$ and radius $R_{wd}$ of the white dwarfs are typically given by $0.5M_{\odot}$
and $10^4$km, respectively. Thus it approximately takes $0.1$ second for the axion stars to pass the
atmospheres of the white dwarfs.
A distinctive feature is that
the radiations from the magnetic white dwarfs with $B\sim 10^9$G
have wider bandwidths than those of the radiations from the neutron stars.
Since the temperatures of the white dwarfts are equal to or less than $10^4$K and the magnetic field
is equal to $10^9$G, we find using eq(\ref{th}) that
the fluctuations $\omega_{\rm th}/\omega$ is three times larger than
those of the radiations from the neutron stars with $T=10^5$K and $B=10^{10}$G.

\vspace{0.1cm}
When we observe them at earth, they receive gravitational and cosmological red shifts. 
The effect of the gravitational red shift
is, however, very small; $\sqrt{1-2GM_{wd}/R_{wd}}\simeq 1$.
Hence, the frequencies observed at the earth are given by $(\omega \pm \omega_{\rm th})(1+z)^{-1}$, which are
larger than the frequencies of the radiations 
from neutron stars located at the places with the redshift $z$.

\vspace{0.1cm}
We should mention that the radio bursts from white dwarfs with $B=10^9$G are more energetic
than those of the observed FRBs. This is because the whole energies of the axion stars 
colliding with such white dwarfs
are transformed into radiations. The energies are of the order of  
$10^{-12}M_{\odot}\simeq 10^{43}$erg. As we have shown, the axion stars are much smaller than
the white dwarfs so that the whole of axion stars collide with the white dwarfs.

Actually, the emission rate of the radiation energy produced by a single electron in the atmospheres is
given by

\begin{equation}
\dot{w}\equiv \frac{2e^2\dot{p}^2}{3 m_e^2}
\simeq 0.7\times 10^{-11}\mbox{GeV/s}\Big(\frac{10^2\rm km}{R_a}\Big)^4\Big(\frac{10^{-5}\rm eV}{m_a}\Big)^2
\Big(\frac{B}{10^{9}\mbox{G}}\Big)^2,
\end{equation}
where the radiations are also dipole ones.

The electrons in a volume $\lambda^3$ 
( $\lambda=2\pi/m_a\simeq 10\mbox{cm}(10^{-5}\mbox{eV}/m_a)$ denotes the wave length of the radiations )
coherently emit the radiations. 
Thus, the emission rate of the coherent radiations in the volume is

\begin{equation}
\dot{w}(n_e\lambda^3)^2\simeq 10^{39}\mbox{GeV/s}\Bigl(\frac{B}{10^9G}\Bigr)^2
\Bigl(\frac{n_e}{10^{22}\mbox{cm}^3}\Bigr)^2
\end{equation}
where $n_e$ denotes number density of electrons in the atmospheres.
On the other hand, the axion stars have the energies $10^{-12}M_{\odot}(\lambda/R_a)^3
\simeq 10^{33}$GeV
in the volume $\lambda^3$. Thus, when the axion stars pass the atmospheres in a period $0.1$second,
their whole energies are transformed into the radiations.
In this way, when the axion stars collide with the magnetic white dwarfs,
they disappear emitting the radiations.

It is easy to show that the atmospheres of the white dwarfs are transparent for the radiations.
In the formulae given above, taking the parameters $B=10^9$G and $T=10^4$K,
we find that the optical depth $\tau_{\pm}(r_c)=\int_{r=r_c}^{\infty}dr'C_{\pm}(r')<1$
even at $r=r_c$ in which $n(r_c)=10^{22}$cm$^{-3}$.

\vspace{0.1cm}
We should make a comment that the typical white dwarfs have magnetic fields 
$B\sim 10^7$G much smaller than $10^9$G. 
It leads to the numerical parameters $\omega^2/\omega_p^2\simeq 10^{-3}$,
$b\simeq 0.134$ and $\theta\simeq 1/b$. Then, we find that the atmospheres are not transparent when
the radiations are produced in the deep inside of the atmosphere 
with the electron density such as $n_e=10^{22}$cm$^{-3}$.
However, when they are produced at the depth with $n_e=10^{13}$cm$^{-3}$,
they can pass through the atmospheres. But the emission rate in the volume $\lambda^3$ is much small,

\begin{equation}
\dot{w}(n_e\lambda^3)^2\simeq 10^{17}\mbox{GeV/s}\Bigl(\frac{B}{10^7G}\Bigr)^2
\Bigl(\frac{n_e}{10^{13}\mbox{cm}^3}\Bigr)^2.
\end{equation}

Thus, the axion stars emit radiations with their energies $\dot{w}(n_e\lambda^3)^2(R_a/\lambda)^3
\sim 10^{35}$GeV/s. 
Hence, the collisions between the white dwarfs with $B\sim 10^7$G and the axion stars 
does not produce the radiations with enough luminosities to be observed at the earth
when they arise in extragalactic origins.

\vspace{0.1cm}
If the number of white dwarfs in a typical galaxy is of the order of $10^{12}$,
only a small fraction $10^{-4}\sim 10^{-5}$ of the white dwarfs would be those  
with strong magnetic fields $\ge 10^9$G and hydrogen atmospheres.
Then,
the production rate $R_{burst}$ of the FRBs emitted 
in the collisions with such magnetic white dwarfs
is found such that 
$R_{burst}\sim (10^{-2}$/year $\sim 10^{-3}$/year) in a galaxy.
The values are obtained by using the formula in eq(\ref{Rb}) with the use of $R_{wd}=10^4$km.
The rate is ten times larger than or equal to the rate of the FRBs actually observed.
In other words, if the typical number of such magnetic white dwarfs in a galaxy
is of the order of $10^7$, 
the rate of the bursts is approximately equal to the rate of the FRBs observed.
Thus, the FRBs associated with the white dwarfs can be observed with
their frequencies in a range $2$GHz $\sim 3$GHz.
Obviously, they can be distinguished from those arising from the collisions with neutron stars.

The number of the white dwarfs with strong magnetic fields $\ge 10^9$G in a galaxy is not known
and the estimation of the number is difficult. Although we know the presence of such white dwarfs,
the number of them could be very few. Thus, the event rate of the FRBs associated 
with the white dwarfs\cite{wd}
could be much small so that the FRBs are undetectable. 


\section{summary and discussions}
\label{9}
We have shown the details of a possible production mechanism of FRBs; 
FRBs arise from the collisions between axion stars and neutron stars.
We have found that the masses and radii of the axion stars are 
given by $M_a\sim 10^{-12}M_{\odot}$ and $R_a\sim 10^2$km, respectively.
The axion stars are rapidly converted into radiations under strong magnetic fields
of the neutron stars.
The radiations are emitted in the atmospheres of the neutron stars.
We have shown that the atmospheres are transparent for the radiations.
The transparency comes from the presence of strong magnetic fields $B\ge 10^{10}$G
and the low frequencies $\sim 1$GHz of the bursts.
According to the mechanism,
we can explain naturally the durations ( $\sim $ ms ) and amount of the energies ( $10^{40}$erg )
of the bursts.

It apparently seems that the radiations is monochromatic with the frequency given by the
axion mass. But, the observed frequencies
have finite bandwidths. We have shown that
the bandwidths are caused by the thermal fluctuations of electrons
emitting the radiations.
   
In the actual collisions the tidal forces of the neutron stars distort the formation
of the axion stars. When the axion stars are close to the neutron stars,
the gravitational forces of the neutron stars 
are stronger than those of axion stars binding themselves. 
Then, the axions freely fall to the neutron stars.
But the coherence of the axions is kept because the number density of
the axions in the volume $m_a^{-3}$ is quite large.

Our mechanism predicts that there are no radiations with any frequencies after the bursts.
This is consistent with the results of follow-up observations\cite{frb3}.  
It also predicts that FRBs contain circular polarizations.
The circular polarizations arise owing to the absorption
of either right or left handed polarized radiations
in the magnetospheres.
Circular polarizations have recently been observed\cite{frb3} in a FRB. 

Similar radio bursts may arise when the axion stars collide with magnetic white dwarfs. 
We have found that the duration of the bursts is of the order of $0.1$second
and that the radiations have wider bandwidths than those of the radiations from neutron stars. 
The features can be observable 
only if the white dwarfs have strong magnetic fields $\ge 10^9$G.
Although the number of such white dwarfs in a galaxy is unknown,
the production rate of the bursts is sufficienlly large for them to be detectable
if their number is larger than $10^6$ in a galaxy.

\vspace{0.1cm}
If the our production mechanism of FRBs is true, we can reach a significant conclusion 
that the axions are the dominant component of dark matter
and their mass is about $10^{-5}$eV, which is
in the window allowed by observational and cosmological
constraints\cite{axionc}.

 \vspace{0.2cm}
The author
expresses thanks to Prof. J. Arafune for useful comments
and discussions.



\end{document}